%% file: main.tex
\documentclass{article}

\usepackage[utf8]{inputenc}
\usepackage{microtype}
\usepackage{graphicx}
\usepackage{subfigure}
\usepackage{booktabs} % for professional tables
\usepackage[normalem]{ulem}
\usepackage{xspace}
\usepackage{relsize}
\usepackage{soul}
\usepackage{makecell}
\usepackage{hyphenat}
\usepackage{multirow}
\usepackage{amssymb}
\usepackage{amsmath}
\usepackage{dblfloatfix}
\usepackage{float}
\usepackage{tabularx}
\usepackage{seqsplit}
\usepackage{physics2}\usephysicsmodule{ab}
\usepackage{enumitem}
\usepackage{comment}
\usepackage{xcolor}
\usepackage{import}
\usepackage{svg}
\usepackage{placeins}
\usepackage{tikz}\usetikzlibrary{arrows.meta,positioning,decorations}
\usepackage{pgfplots}\pgfplotsset{compat=1.18}
\usepackage{capt-of}  %
\usepackage{cuted}    %
\usepackage{algorithm}
\usepackage{algpseudocode}
\usepackage{threeparttable}
\usepackage{titlesec}
\usepackage{pifont}
\usepackage{csquotes}
\usepackage{romannum}
\usepackage[group-separator={,}]{siunitx}

\newcommand{\sysname}{PROBE}

\usepackage{hyperref}

\usepackage[accepted]{mlsys2025}

\mlsystitlerunning{PROBE: Co-Balancing Computation and Communication in MoE Inference via Real-Time Predictive Prefetching}

\begin{document}

\twocolumn[
    \mlsystitle{PROBE: Co-Balancing Computation and Communication in MoE Inference via Real-Time Predictive Prefetching}

    % unless you've provided the [accepted] option to the mlsys2025

    % Academic affiliations should list Department, University, City,
    % Region, Country

    % \mlsyssetsymbol{equal}{*}

    \begin{mlsysauthorlist}
        \mlsysauthor{Qianchao Zhu}{to}
        \mlsysauthor{Xucheng Ye}{to}
        \mlsysauthor{Yuliang Liu}{to}
        \mlsysauthor{Haodong Ouyang}{to}
        \mlsysauthor{Chengru Song}{to}
    \end{mlsysauthorlist}

    \mlsysaffiliation{to}{Kling Infra, Kuaishou Technology}

    \mlsyscorrespondingauthor{}{}

    \mlsyskeywords{Machine Learning, MLSys}

    \vskip 0.3in
    \vspace{-1em}
    \begin{abstract}
    Mixture-of-Experts models have become a dominant architecture for scaling Large Language Models by activating only a sparse subset of experts per token. However, latency-critical MoE inference faces a fundamental tension: while expert parallelism improves memory efficiency, it also amplifies execution stragglers. In real-world serving, continuous batching and diverse concurrent requests induce rapid semantic shifts, causing expert hotspots to migrate abruptly across GPUs and triggering the ``double penalty'' of coupled computational skew and network congestion.

    We propose \textbf{\sysname{}}, an inference system that co-balances computation and communication in real time. \sysname{} introduces \textbf{Continuous Lookahead Pipelining}, which proactively predicts, plans, and prefetches for upcoming layers while keeping all control overheads off the critical path. \sysname{} consists of: (1) a \textit{Gate-Initialized Lookahead Predictor} that distills the target router to forecast next-layer expert activation with high fidelity; (2) a \textit{Hardware-Aware Balance Planning} solver that jointly optimizes dynamic expert replication and token assignment under strict hiding-window constraints; and (3) a \textit{Phase-Locked Co-Scheduling} policy that uses split-phase transmission to hide bandwidth-intensive expert transfers behind computation without contending with All-to-All collectives. Experiments show that \sysname{} reduces prefill latency by up to \textbf{1.32$\times$} and improves decoding throughput by up to \textbf{1.26$\times$} over state-of-the-art baselines, especially under extreme workload volatility.
    %  actually on NVIDIA H200

\end{abstract}
]


% The command takes one argument, which is text to display at the
% start of the footnote.

\printAffiliationsAndNotice{}  % leave blank if no need to mention
% equal contribution
% \printAffiliationsAndNotice{\mlsysEqualContribution} % otherwise use
% the standard text.

\input{sections/1-introduction}

\input{sections/2-background}
\input{sections/3-modeling}
\input{sections/4-method}

\input{sections/5-implementation}
\input{sections/6-experiment}

\input{sections/7-conclusion}


\clearpage
\bibliography{main}
\bibliographystyle{mlsys2025}

%%%%%%%%%%%%%%%%%%%%%%%%%%%%%%%%%%%%%%%%%%%%%%%%%%%%%%%%%%%%%%%%%%%%%%%%%%%%%%%
%%%%%%%%%%%%%%%%%%%%%%%%%%%%%%%%%%%%%%%%%%%%%%%%%%%%%%%%%%%%%%%%%%%%%%%%%%%%%%%
% SUPPLEMENTAL CONTENT AS APPENDIX AFTER REFERENCES
%%%%%%%%%%%%%%%%%%%%%%%%%%%%%%%%%%%%%%%%%%%%%%%%%%%%%%%%%%%%%%%%%%%%%%%%%%%%%%%
%%%%%%%%%%%%%%%%%%%%%%%%%%%%%%%%%%%%%%%%%%%%%%%%%%%%%%%%%%%%%%%%%%%%%%%%%%%%%%%
\clearpage
% \appendix
% \input{sections/appendix}

%%%%%%%%%%%%%%%%%%%%%%%%%%%%%%%%%%%%%%%%%%%%%%%%%%%%%%%%%%%%%%%%%%%%%%%%%%%%%%%
%%%%%%%%%%%%%%%%%%%%%%%%%%%%%%%%%%%%%%%%%%%%%%%%%%%%%%%%%%%%%%%%%%%%%%%%%%%%%%%

\end{document}

%% file: sections/1-introduction.tex
\section{Introduction}
\label{sec:introduction}

\begin{figure}[t]
    \centering
    \includegraphics[
        width=0.85\linewidth
    ]{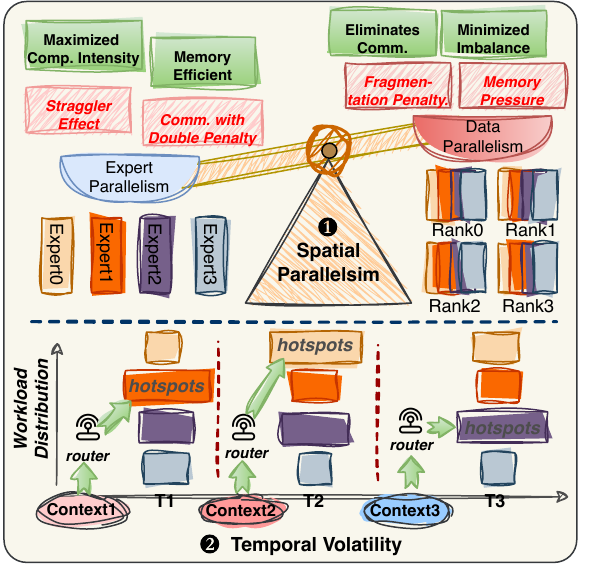}
     \caption{\textbf{Spatial and Temporal Challenges in MoE Inference.} Under Expert Parallelism, efficiency is constrained by: \textbf{(1) Spatial imbalance}, where skewed token-to-expert routing creates computational stragglers and communication bottlenecks; and \textbf{(2) Temporal volatility}, where expert hotspots shift rapidly over time under continuous batching. This motivates a system that can handle both instantaneous skew and continuous distribution shifts.}
     \vspace{-2em}
    \label{fig:triangle}
\end{figure}

The insatiable demand for model intelligence has driven Large Language Models (LLMs)~\cite{vaswani2017attention,comanici2025gemini,achiam2023gpt,liu2025deepseek} toward trillion-parameter scales~\cite{kaplan2020scaling}. To sustain this scaling without incurring prohibitive computational costs, the Mixture-of-Experts (MoE) architecture has emerged as the de facto standard~\cite{fedus2022switch,jiang2024mixtral,du2022glam}. By decoupling parameter count from active computation—activating only a sparse subset of experts per token—MoE enables models like GPT-OSS~\cite{agarwal2025gpt}, DeepSeek-V3~\cite{liu2024deepseek}, Qwen3-MoE~\cite{yang2025qwen3} to achieve massive capacity with manageable FLOPs.

While expert load balancing has been extensively explored in the training regime—often relying on auxiliary losses~\cite{lepikhin2020gshard,fedus2022switch} or capacity constraints~\cite{he2022fastermoe,zhai2023smartmoe}, recent state-of-the-art models have increasingly shifted toward finer-grained sparsity~\cite{team2025kimi} and deep expert specialization~\cite{liu2025deepseek,yang2025qwen3,wang2024auxiliary} to enhance model capabilities. This paradigm shift relaxes balancing constraints and significantly intensifies workload skewness. Consequently, serving these models for latency-critical inference creates a fundamental tension between maintaining memory efficiency, mitigating severe stragglers, and handling dynamic imbalance.

As illustrated in \autoref{fig:triangle}, while Expert Parallelism (EP) enables massive models to fit within GPU memory, the resulting workload imbalance becomes a critical bottleneck for inference efficiency. This performance degradation is driven by a complex interplay of spatial and temporal dimensions.
\textbf{Spatially}, unlike the uniform workload of dense models, MoE routing and execution is dictated by input semantics; popular experts create ``hotspots" that inflict a \textit{double penalty}, where the overloaded rank is simultaneously throttled by computational skew and network congestion during All-to-All collectives.
\textbf{Temporally}, this instability is exacerbated by the stochastic nature of continuous batching~\cite{kwon2023efficient}, where the global batch composition churns rapidly as requests join and depart at arbitrary intervals. Consequently, expert hotspots migrate abruptly, especially during the prefill phase with limited steps, rendering static expert placement obsolete.

To mitigate the straggler effect, existing solutions~\cite{li2025speculative,han2025grace,yun2024flex,he2022fastermoe,dai2024deepseekmoe, doucet2025harmoeny,zeng2025efficientmoe} largely converge on selective expert replication, trading memory for improved load balance. However, this paradigm fails to adequately address the unique challenges of latency-critical inference.
First, strategies that permanently replicate popular experts inflate memory usage, competing with the KV cache for limited HBM capacity. Second, expert hotspots exhibit high-frequency shifts, a volatility that is particularly acute during the condensed prefill phase.
Consequently, reactive approaches relying on historical statistics inherently lag behind abrupt variations, producing obsolete placement decisions.
Furthermore, given the strict requirement to avoid stalling the critical path, any exposed overhead (e.g., offloading-based approaches~\cite{hu2026brownoutserve,yu2026taming}, host-based balancing solvers, or blocking expert transfers) can neutralize the balancing gains. Finally, techniques relying on pre-gated routing~\cite{hwang2024pre} require training-time adaptation, violating strict correctness requirements.

In this paper, we propose \textbf{\sysname{}}, an inference system designed to co-balance computation and communication in real-time. \sysname{} fundamentally shifts the paradigm from \textit{reactive adjustment} to \textit{proactive preparation}. While token arrival is stochastic, the semantic routing of deep models is predictable.
\sysname{} implements a \textbf{Continuous Lookahead Pipelining} mechanism. 
Instead of blocking the critical path, \sysname{} overlaps the \textit{Predict}, \textit{Plan}, and \textit{Prefetch} phases for the subsequent layer with the main stream, effectively hiding these control overheads.

Specifically, we make the following contributions:

\textbullet \enspace \textbf{\underline{Gate-Initialized Lookahead Predictor:}} We introduce a lightweight predictor that distills the routing logic of the target layer. By freezing the target layer's router as a prior and leveraging the previous layer's hidden states as input, it forecasts next-layer expert hotspots with $\approx90\%$ accuracy while incurring negligible overhead.

\textbullet \enspace \textbf{\underline{Hardware-Aware Balance Planning:}} We formulate straggler mitigation as a resource assignment problem. Unlike solvers that ignore transfer costs, our planner strictly bounds expert replication decisions within the device-specific ``hiding window'', dynamically replicates experts onto underutilized ranks, and ensures these overheads do not stall the pipeline.

\textbullet \enspace \textbf{\underline{Phase-Locked Co-Scheduling:}} \sysname{} operates on a dual-track architecture. Within this framework, we design a split-phase transmission scheduling that orchestrates prediction, planning, and prefetching to execute orthogonally to the main stream. This ensures that bandwidth-heavy prefetching never contends with All-to-All collectives, guaranteeing zero contention on hardware resources.

\textbullet \enspace \textbf{\underline{Evaluation:}} Experiments show that \sysname{} effectively neutralizes stragglers, reducing prefill latency by up to \textbf{1.32$\times$} and improving decoding throughput by up to \textbf{1.26$\times$} over state-of-the-art baselines under extreme workload volatility. 
%  on NVIDIA H200

%% file: sections/2-background.tex
\section{Background}
\label{sec:background}

The MoE architecture has emerged as a dominant paradigm for scaling LLMs by decoupling parameter size from computational cost~\cite{jacobs1991adaptive, shazeer2017outrageously, fedus2022switch}. While historical approaches addressed expert load balancing during training via auxiliary losses~\cite{lepikhin2020gshard,xue2024openmoe}, recent state-of-the-art models increasingly prioritize expert specialization and adopt finer-grained expert granularity with higher sparsity to maximize performance~\cite{liu2025deepseek,guo2025advancing,qiu2025demons,team2025kimi}, often relaxing balancing constraints. 
This shift significantly intensifies workload skewness during inference: under expert parallelism, the synchronous execution of MoE layers transforms skew into a severe \textit{straggler effect}, bounding layer latency by the most heavily loaded device.
In this section, we (\S~\ref{subsec:characterizing_imbalance}) characterize the manifestations of load imbalance in real-time inference, (\S~\ref{subsec:dilemma}) analyze the trade-offs between parallelism strategies, and (\S~\ref{subsec:challenges}) identify the system constraints imposed by low-latency serving.

\subsection{Characterizing Expert Load Imbalance}
\label{subsec:characterizing_imbalance}

\begin{figure}[t]
    \centering
    \includegraphics[
        width=\linewidth
    ]{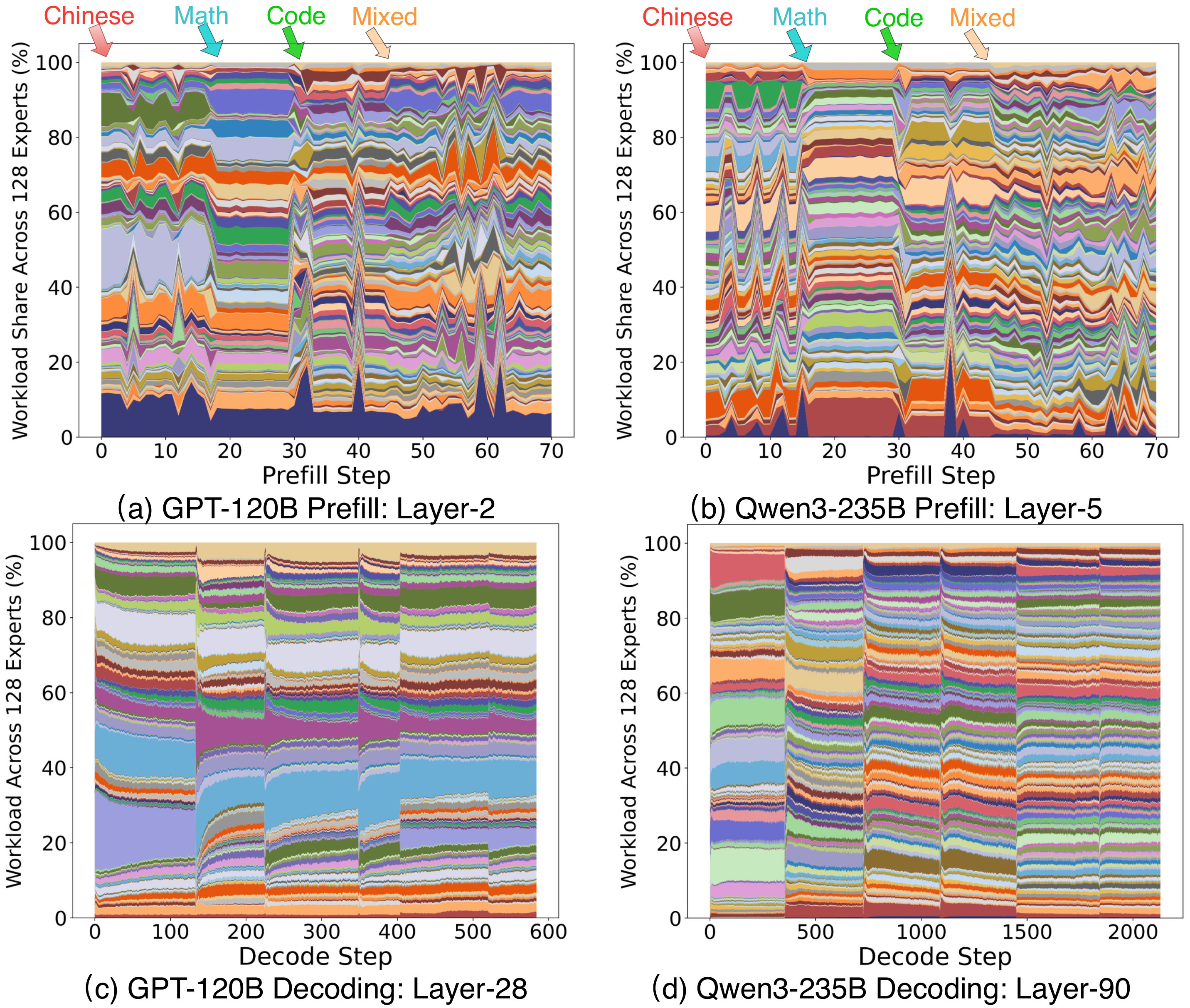}
    \caption{\textbf{Expert activation patterns across prefill and decoding.} Measurements use $ep=8$ with a standard sharded expert placement policy.
    Subfigures (a) and (b) show the concentrated, bursty skew during prefill ($\approx$32K tokens).
    Subfigures (c) and (d) show rapid load shifts during decoding ($\approx$8K tokens), where expert popularity changes with semantic transitions.
    Comparing GPT-OSS-120B (Top-4 out of 128 experts) and Qwen3-235B (Top-8 out of 128 experts) illustrates that model sparsity patterns further modulate imbalance severity.}
    \vspace{-1em}
    \label{fig:imbalance}
\end{figure}

To systematically analyze the straggler effect in EP, we quantify load skewness using the \textit{Imbalance Ratio} ($\mathcal{IR}$). Defined at the rank granularity, $\mathcal{IR}$ measures the disparity between the maximum and average workload across an EP group of size $ep$. For a rank $r$ hosting experts $\mathcal{E}_r$, the local load is $\mathcal{L}_{r}=\sum_{e \in \mathcal{E}_r} n_{e}$, yielding:
\begin{equation}
\mathcal{IR} = \frac{\max_{r \in \{0, \dots, ep-1\}} \mathcal{L}_{r}}{\frac{1}{ep} \sum_{r=0}^{ep-1} \mathcal{L}_{r}}
\end{equation}
An $\mathcal{IR}$ of 1.0 represents ideal balance. However, as $ep$ scales MoE models with higher sparsity, the probability of ``hot'' experts colliding on a single device increases.
Consequently, a substantially elevated $\mathcal{IR}$ implies that the cluster's aggregate throughput is throttled by the straggler device, forcing underutilized resources to idle at synchronization barriers.

\paragraph{Prefill: Burstiness from Semantic Clustering.}
During prefill, the parallel processing of massive prompt sequences triggers severe workload concentration. Unlike the uniform distribution assumed by statistical multiplexing, the semantic locality inherent in input contexts causes specific experts to be disproportionately activated. 
As shown in Figure~\ref{fig:imbalance}(a-b), injecting new datasets transforms prompt semantics into instantaneous traffic bursts, manifesting as frequent spikes in the $\mathcal{IR}$ above \textbf{2.6} even with large batches ($\approx$32K tokens).
Consequently, the overloaded rank dictates global tail latency, significantly inflating Time-To-First-Token (TTFT).

\paragraph{Decoding: Volatility under Continuous Batching.}
In contrast to the concentrated skewness observed in prefill, the decoding phase is characterized by rapid load volatility. 
While token aggregation from diverse semantics results in a lower peak $\mathcal{IR}$, the workload distribution is destabilized by the mechanics of continuous batching~\cite{kwon2023efficient}.
The constant churn of arriving and departing requests, coupled with the semantic evolution of generated sequences, results in an unstable expert distribution.
As shown in Figure~\ref{fig:imbalance}(c-d), expert popularity shifts unpredictably during workload transitions, causing $\mathcal{IR}$ to fluctuate between \textbf{1.43} and \textbf{2.28}.
Crucially, this volatility creates a significant bottleneck, forcing approximately 50\% of global compute capacity to idle at synchronization barriers.

\subsection{The Dilemma of MoE Parallelism} 
\label{subsec:dilemma}

% Figure for Computation Latency

To scale the inference of massive MoE models, existing frameworks have largely converged on a hybrid paradigm: applying Data Parallelism (DP) to attention modules and EP to MoE modules~\cite{zheng2024sglang, kwon2023efficient, dai2024deepseekmoe, nvidia_tensorrt_llm}. While EP is essential for managing the massive parameters of modern MoE models, it introduces a fundamental tension between memory efficiency and straggler effect. EP maximizes batch size per expert but suffers from the straggler effect. Conversely, DP eliminates imbalance but necessitates full weight replication, which is often infeasible. We deconstruct these trade-offs across three critical dimensions:

\paragraph{Memory Efficiency: Capacity vs. Bandwidth.} 
As MoE models scale toward trillions of parameters, the full weight replication required by DP becomes prohibitively expensive, making EP a necessity for memory capacity. Beyond mere storage, EP optimizes memory bandwidth through global token consolidation. By aggregating tokens across all ranks, EP ensures large effective batch sizes for each expert. In contrast, DP fragments the global batch across independent replicas. This fragmentation forces the loading of full expert weights for a small number of local tokens, drastically inflating redundant memory accesses and computation. Consequently, DP degrades arithmetic intensity, pushing the computation into a  memory-bound bottleneck, particularly for ``cold'' experts with negligible utilization.

\begin{figure}[t]
    \centering
    \includegraphics[width=0.95\linewidth]{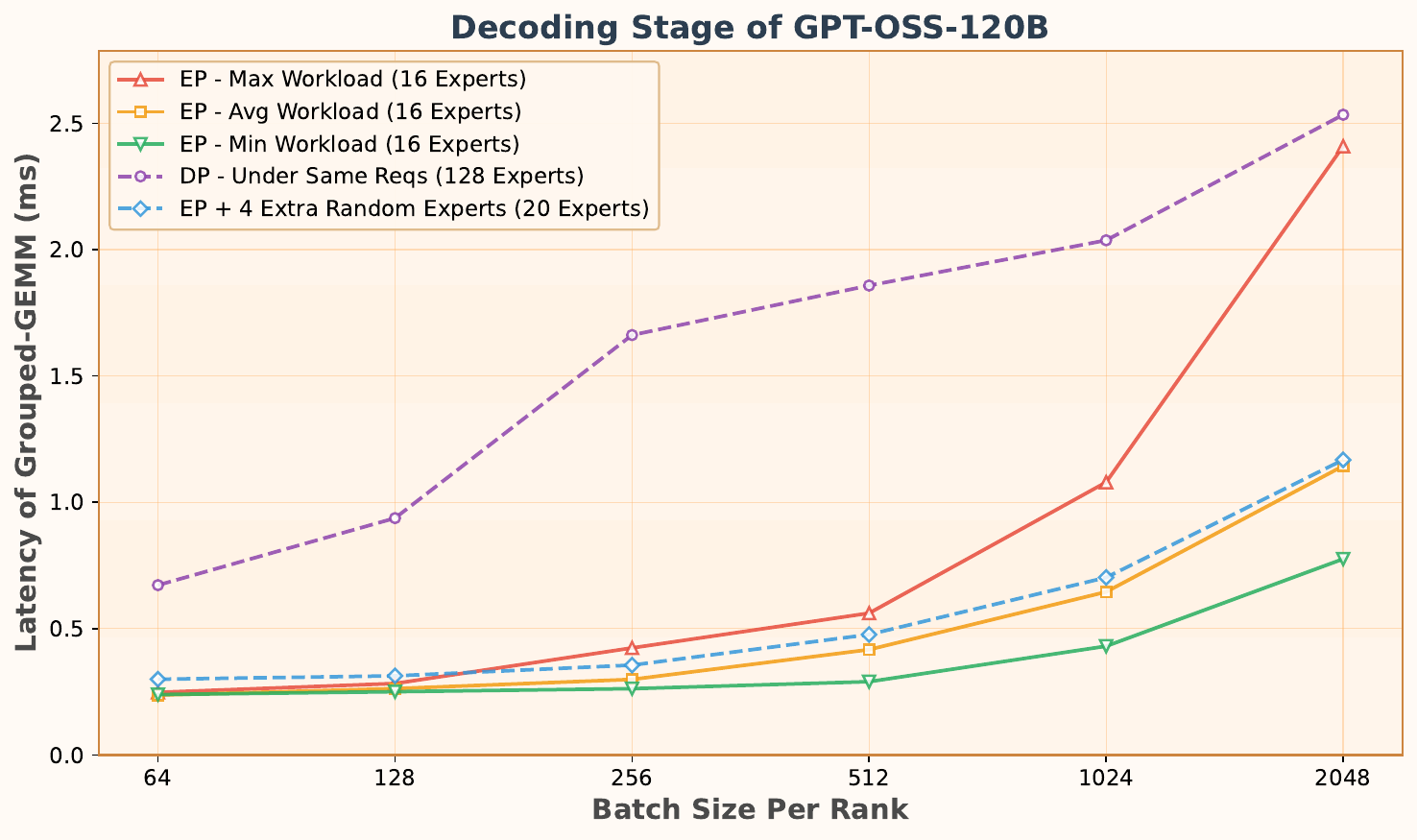}
    \caption{\textbf{MoE compute latency.} Profiling via SGLang on GPT-OSS-120B (128 experts, Top-4). We compare EP (Max/Avg/Min) with DP and EP + 4 extra experts. DP is bottlenecked by fragmentation (low arithmetic intensity and padding), while modest EP redundancy mitigates stragglers with minimal memory overhead.}
    \label{fig:compute-balance}
    \vspace{-1em}
\end{figure}

\paragraph{Computational Intensity:  Straggler \& Fragmentation.}
MoE efficiency hinges on the arithmetic intensity of Grouped GEMMs. EP maximizes this by aggregating tokens to saturate Tensor Cores, yet incurs severe load imbalance. As quantified in Figure~\ref{fig:compute-balance}, the significant gap between maximum and average rank latency confirms that system throughput is bound by the single slowest straggler.
In contrast, DP eliminates imbalance but suffers a severe \textbf{fragmentation penalty}. Processing fragmented local batches dilutes arithmetic intensity, pushing computation into a memory-bound regime.
Additionally, rigid kernel tiling necessitates extensive padding for irregular token counts; the resulting waste in FLOPs becomes increasingly severe with fine-grained sparsity~\cite{guo2025sonicmoe}. 
Crucially, the ``EP + Extra Experts'' profile reveals that the bottleneck is skewness, not aggregate workload. Selectively replicating experts reduces the tail latency with minimal memory overhead, effectively balancing load without incurring the performance penalties of full DP.

\begin{figure}[t]
    \centering
    \includegraphics[width=0.96\linewidth]{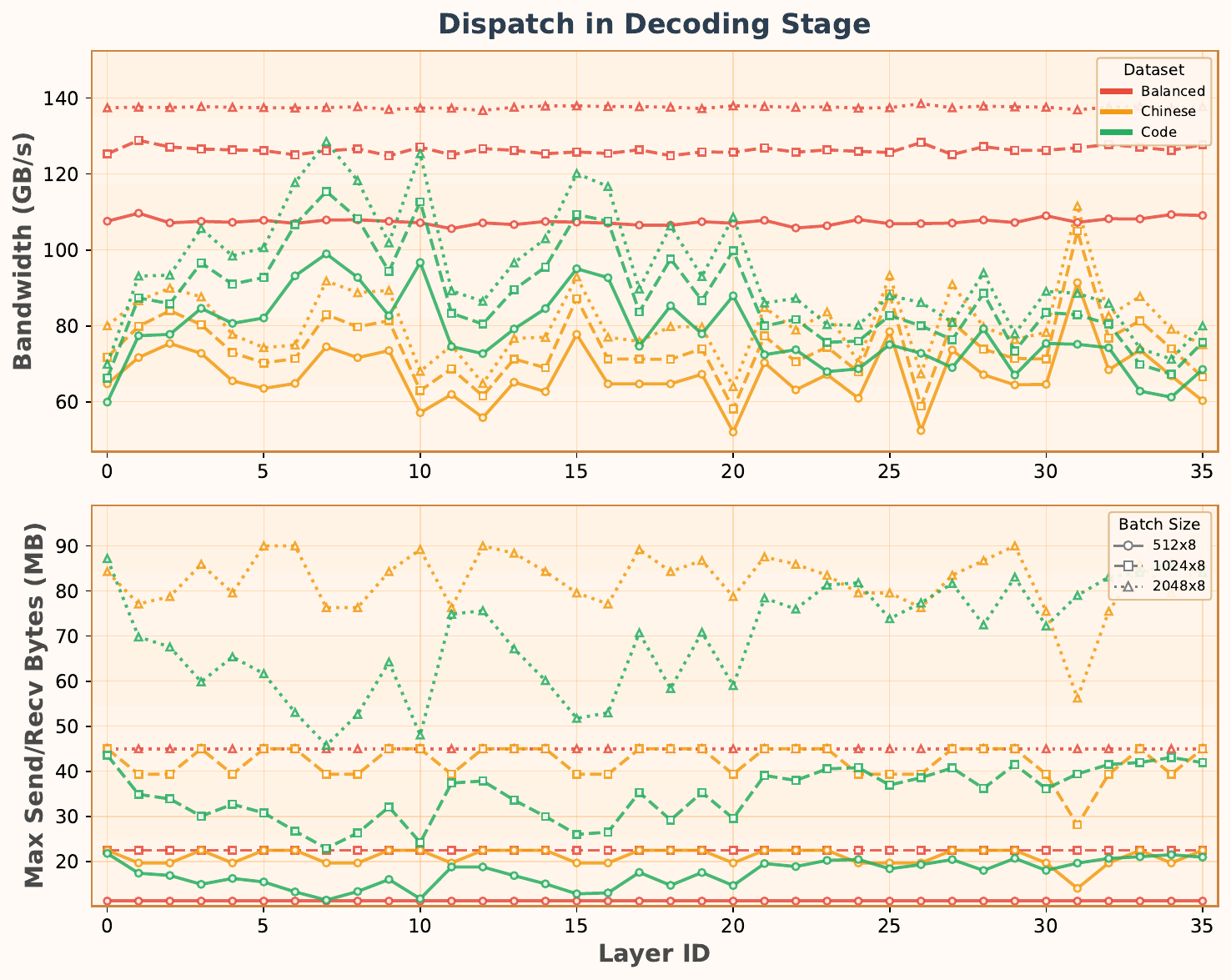}
    \caption{\textbf{Skew hurts All-to-All efficiency.} Benchmarked on  8$\times$H800, GPT-OSS-120B with DeepEP~\cite{zhao2025deepep}. Top: effective All-to-All Dispatch bandwidth. Bottom: max per-rank traffic volume. Compared to a manually balanced top-$K$ baseline, real workloads create receiver hotspots and reduce effective bandwidth; Combine phase shows similar behavior.}
    \label{fig:comm-balance}
    \vspace{-1em}
\end{figure}

\paragraph{Communication: Coupled Skew and the Double Penalty.}
EP relies on bandwidth-bound All-to-All collectives, where workload skew creates a \textbf{double penalty}: computational hotspots are inevitably coupled with network congestion.
Even with topology-aware and token deduplication optimizations~\cite{zhao2025deepep}, ranks hosting popular experts attract excessive unique tokens. 
Figure~\ref{fig:comm-balance} confirms that this skew drastically inflates the maximum receive volume on specific ranks. Since collective operations are synchronized by the slowest device, this local congestion collapses effective cluster-wide bandwidth. Consequently, the overloaded rank suffers as a compounded bottleneck, sequentially throttled by high network ingress, computation, and egress.

\subsection{Challenges in Real-Time MoE Inference}
\label{subsec:challenges}

To navigate the complex trade-offs between memory, computation, and communication, recent research has explored various hybrid strategies. 
Approaches like Grace-MoE~\cite{han2025grace} and Libra employ expert replication to trade memory for better load balance, while systems such as FasterMoE~\cite{he2022fastermoe} and FlexMoE~\cite{yun2024flex} leverage dynamic offloading to smooth out peaks. 
However, transposing the paradigm that predominantly designed for throughput-oriented settings to latency-critical inference is challenging.
Unlike training, where complex solver costs can be amortized over long backward passes, online inference operates under strict Service Level Objectives. In this regime, even modest scheduling overheads or reactive data movement can nullify the performance gains. Consequently, adapting dynamic balancing to online inference requires overcoming three distinct hurdles:

\paragraph{Expert Hotspot Shifts in Inference}
The first challenge stems from the inherently dynamic routing behavior~\cite{zhang2025advancing}. Unlike the static batches in training, inference engines handle requests that join and depart at arbitrary intervals. This stochasticity introduces high variance in token distribution per step, causing rapid shifts in expert popularity that render historical heuristic methods ineffective.
Therefore, the system requires a high-fidelity predictor capable of anticipating router decisions ahead of time. Crucially, this prediction must feed into a lightweight but effective solver that constructs a distribution strategy in real time, accounting for the constantly fluctuating batch composition rather than relying on stale statistics.

\paragraph{Enforcing Zero-Overhead Balancing.}
The stringent Time-Per-Output-Token (TPOT) constraints mandate that auxiliary load-balancing operations remain strictly hidden behind the critical path~\cite{zhao2024blendserve,chen2024centauri}.
This creates a strong dependence on hardware characteristics.
On configurations with high compute capability but limited interconnect bandwidth, fast compute kernels (e.g., Attention or Grouped GEMM) shrink the overlap window available for expert transfers, forcing the system to cap the transfer volume.
Conversely, limited bandwidth can prolong the All-to-All communication phase, inadvertently widening the window available for solver execution and tolerating more planning steps.
This imposes a rigid constraint: the load balancer must be hardware-aware, dynamically trading off solver complexity against transfer volume to maximize balance gains within device-specific execution budgets without stalling the critical pipeline.

\paragraph{Mandating CUDA Graph Compatibility.}
Modern serving engines often rely on CUDA Graph~\cite{nvidia_cuda_guide} to accelerate the decoding phase~\cite{zheng2024sglang,kwon2023efficient}. Without CUDA Graph, CPU-side dispatch overhead for long sequences of kernels prevents back-to-back execution, leaving large idle bubbles on the device timeline.
However, dynamic load balancing creates a fundamental conflict: variable control flow in solvers and dynamic P2P expert transfers can preclude static graph capture, forcing a regression to eager execution.
To preserve graph-based speedups, the balancing mechanism must eliminate host-device synchronization (e.g., host-based ILP solvers). This motivates implementing solvers natively on the GPU, leveraging kernel fusion, and developing custom communication kernels to absorb dynamic logic while adhering to CUDA Graph requirements.

%% file: sections/3-modeling.tex
\section{Performance Modeling}
\label{sec:modeling}

Building on the system challenges identified, we formulate an analytical performance model for the hybrid parallel setting (DP for Attention, EP for MoE). The model quantifies end-to-end layer latency as a function of token distribution, capturing the interaction among three components: (1) \textbf{computation latency}, inflated by load skew and Grouped GEMM fragmentation; (2) \textbf{communication latency}, which degrades under traffic congestion during both dispatch and combine; and (3) \textbf{prefetch overhead}, bounded by the available hardware overlap window.

\subsection{System Setup and Notation}

\begin{table}[t]
\centering
\footnotesize
\renewcommand{\arraystretch}{1.1} % Slightly increase row height
\begin{tabular}{c|c}
\hline
\textbf{Symbol} & \textbf{Definition} \\
\hline
\multicolumn{2}{l}{\textit{\qquad  System \& Model Parameters}} \\
$ep$ & Expert Parallelism size (number of ranks) \\
$B$ & Global batch size (tokens per step) \\
$H$ & Hidden dimension size \\
$\mathcal{W}$ & Parameter size per expert \\
$\bar{F}$ & Per-token FLOPs per expert \\
$\eta_g(\cdot)$ & GEMM efficiency function (w.r.t. tokens/expert) \\

\hline
\multicolumn{2}{l}{\textit{\qquad   Workload \& Distribution}} \\
$\mathcal{E}_r$  & Set of experts \textit{physically hosted} on rank $r$ \\
$\Delta_r$ & Set of \textit{redundant} experts replicated on rank $r$ \\
$n_e$ & Global tokens routed to expert $e$ ($\sum n_e = B \cdot k$) \\
$n^{r_s}_{e,r_t}$ & Tokens on source $r_s$ routed to  $e$ hosted on  $r_t$ \\
$\lambda^{in/out}_r$ & Token ingress/egress deduplication ratio on rank $r$\\
$\mathcal{V}^{in/out}_r$ & Network traffic volume for rank $r$ \\
$\mathcal{IR}$ & Imbalance Ratio ($\max_r \text{Load}_r / \text{Avg Load}$) \\ 
\hline
\end{tabular}
\caption{Key notation for the MoE performance model.}
\vspace{-1em}
\label{tab:notation}
\end{table}

Table~\ref{tab:notation} summarizes the notation. The router produces a global token count $n_e$ for each expert $e$.
We denote by $n^{r_s}_{e,r_t}$ the number of tokens from source rank $r_s$ routed to expert $e$ on rank $r_t$ (so $\sum_{r_s,r_t} n^{r_s}_{e,r_t} = n_e$). When the source rank is irrelevant, we write $n_{e,r_t} = \sum_{r_s} n^{r_s}_{e,r_t}$.

In a standard EP system with sharded placement (no replication), each expert is hosted on a unique rank, so all tokens of expert $e$ are processed on its home rank.
In the presence of expert replication, the workload $n_e$ can be partitioned across multiple ranks that host a copy of expert $e$; we denote by $n_{e,r}$ the number of tokens of expert $e$ assigned to (and processed on) rank $r$, subject to the conservation constraint $\sum_{r} n_{e,r} = n_e$.

\subsection{Computation: Skew and Fragmentation}

\paragraph{Rank-Level Latency.}
After dispatch, rank $r$ executes its assigned experts using Grouped GEMM. The effective compute time depends not only on FLOPs but also on kernel efficiency $\eta_g(\cdot)$, which degrades for small token counts due to padding and reduced arithmetic intensity.
The processing time for each expert $e$ on rank $r$ is modeled as:
\begin{equation}
T_{e,r}(n_{e,r}) = \frac{n_{e,r} \cdot \bar{F}}{\eta_g(n_{e,r}) \cdot F_{peak}},
\end{equation}
where $\bar{F}$ denotes the per-token FLOPs. The total compute latency for rank $r$ is the summation over all locally hosted experts, comprising both native and replicated: $T^{r}_{comp} = \sum_{e \in \mathcal{E}_r \cup \Delta_r} T_{e,r}$.

\paragraph{Straggler Effect.}
Since EP inference is synchronous, the layer latency is dictated by the slowest rank. We relate the tail latency to the cluster average via the $\mathcal{IR}$:
\begin{equation}
T_{comp} = \max_{r} T^{r}_{comp} \approx \mathcal{IR} \cdot \left( \frac{1}{ep} \sum_{r} T^{r}_{comp} \right).
\end{equation}
This formulation highlights a compounding degradation: high skew ($\mathcal{IR} \gg 1$) creates a straggler rank with large $T^r_{comp}$, while fragmentation of splitting $n_e$ across ranks reduces $n_{e,r}$, pushing computations into the low-efficiency regime of $\eta_g(\cdot)$.

\subsection{Communication: The Double Penalty}

\paragraph{Traffic Volume and Send/Recv Congestion.}
The All-to-All dispatch and combine phases are bandwidth-intensive. Their latency is dictated by the bottleneck rank handling the maximum data volume (either send or receive).
Crucially, token deduplication dynamics differ for ingress and egress traffic.
Let $\lambda^{in}_r$ denote the deduplication factor for traffic received by rank $r$ (i.e., how many experts on $r$ are hit by a unique remote token), and $\lambda^{out}_r$ the corresponding factor for traffic sent out of rank $r$.
The ingress ($\mathcal{V}^{in}_r$) and egress ($\mathcal{V}^{out}_r$) volumes are formulated as:
\begin{equation}
% \mathcal{V}^{in}_r = \frac{H}{\lambda^{in}_r} \sum_{r' \neq r} \sum_{e \in \mathcal{E}_r \cup \Delta_r} n_{e,r'}, \quad 
% \mathcal{V}^{out}_r = \frac{H}{\lambda^{out}_r} \sum_{e \notin \mathcal{E}_r \cup \Delta_r} n_{e,r}. 
\mathcal{V}^{in}_r = \frac{H}{\lambda^{in}_r} \sum_{r' \neq r} \sum_{e \in \mathcal{E}_r \cup \Delta_r} n^{r'}_{e,r}, \quad 
\mathcal{V}^{out}_r = \frac{H}{\lambda^{out}_r} \sum_{e \notin \mathcal{E}_r \cup \Delta_r} n^{r}_{e}.
\end{equation}
The critical communication volume for rank $r$ is determined by the maximum congestion across both directions: $\mathcal{V}_r = \max(\mathcal{V}^{in}_r, \mathcal{V}^{out}_r)$.

\paragraph{Coupled Latency and The Double Penalty.}
Unlike the uniform traffic in DP, EP exhibits a structural correlation between traffic and computation, imposing a double penalty on the straggler rank.
Specifically, the rank $r^*$ hosting global hotspots (maximizing $T^r_{comp}$) inherently attracts the highest volume of unique tokens during the dispatch phase (i.e., $\mathcal{V}^{in}_{r^*} \approx \max_r \mathcal{V}^{in}_r$).
Symmetrically, during the Combine phase, this same rank must redistribute the largest volume of results, creating an egress bottleneck.
Consequently, the end-to-end MoE latency is dictated by this single overloaded device, sequentially throttled by network ingress, computation, and network egress:
\begin{equation}
T_{MoE} \approx \underbrace{\max_{r} T^{r}_{comp}}_{\text{Compute Skew}} + \underbrace{2 \cdot \max_{r} \left( \frac{\mathcal{V}_r}{BW_{net}} \right)}_{\text{Network Skew}}.
\end{equation}
This formulation highlights that the load imbalance $\mathcal{IR}$ proxies the total system slowdown, magnifying the latency penalty beyond computational delays.

\subsection{Constrained Expert Prefetching}

\paragraph{Expert Transfer Latency.}
Dynamic redundancy introduces overheads for moving expert weights. For a rank $r$ prefetching a set of experts $\Delta^{in}_r$ and evicting $\Delta^{out}_r$, the transfer latency is dictated by the maximum of read/write volumes:
\begin{equation}
T^{r}_{trans} = \frac{\max(|\Delta^{in}_r|, |\Delta^{out}_r|) \cdot \mathcal{W}}{BW_{net}}.
\end{equation}

\paragraph{Hiding Window.}
To ensure non-blocking execution on the critical path, expert transfers must be confined within the computation window $T^{r}_{window}$ of non-communication kernels. The exposed overhead is modeled as
$\max\left(0,\; \max_r T^{r}_{trans} - T^{r}_{window}\right)$.
This imposes a hardware-aware constraint: to maintain zero overhead, the system must bound the replica volume $|\Delta_r|$ according to the device's compute-to-bandwidth ratio.

%% file: sections/4-method.tex
\section{System Design}
\label{sec:design}
 
Building on the performance modeling in \S\ref{sec:modeling}, we present \sysname{}, a runtime scheduling system designed to neutralize the straggler effect in latency-critical MoE inference.
\sysname{} introduces a \textbf{Continuous Lookahead Pipelining} paradigm: rather than blocking the critical path to deliberate on load balancing, \sysname{} exploits the execution time of the current layer to predict, plan, and prefetch resources for the next. By decoupling control plane decisions from the main execution flow, \sysname{} achieves just-in-time expert reconfiguration with nearly zero-overhead.

\subsection{Architecture Overview}
\label{subsec:overview}

\begin{figure}[t]
    \centering
    \includegraphics[width=0.95\linewidth]{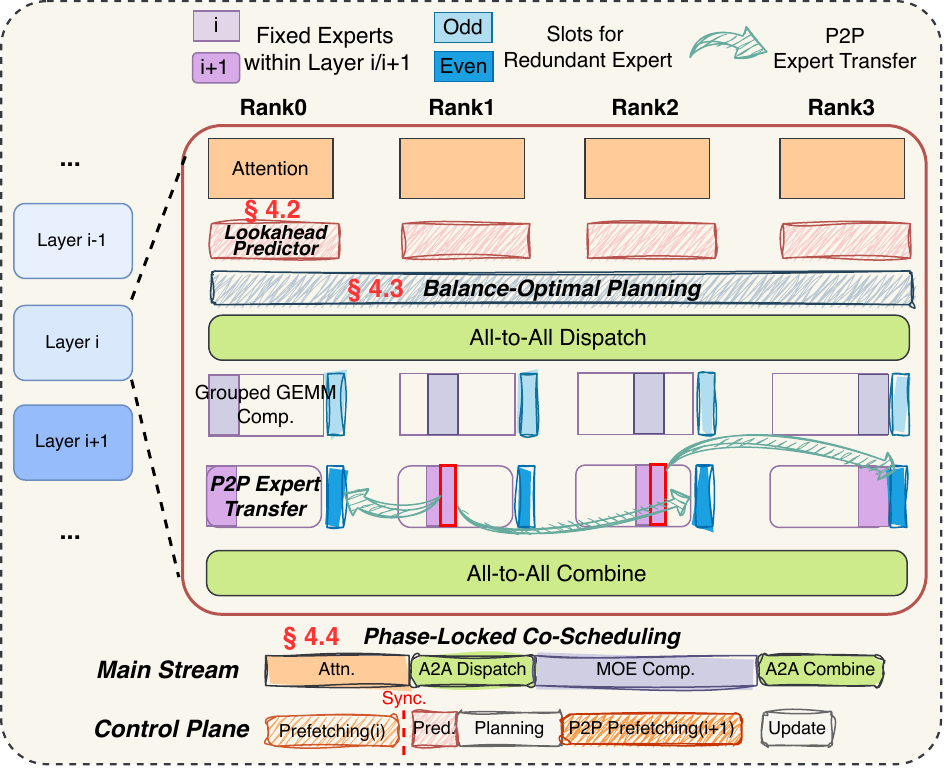}
     \vspace{-1em}
       \caption{\textbf{Overview of \sysname{}.} The system implements a dual-track execution that overlaps control-plane operations with the main stream on complementary resources: predictor/planner run during network-bound All-to-All, while P2P prefetching is overlapped with compute-bound GEMM and attention to hidden management overhead.}
       \vspace{-1em}
    \label{fig:overview}
\end{figure}

As illustrated in Figure~\ref{fig:overview}, \sysname{} establishes a dual-track execution model to isolate management overheads from the critical path:

\begin{itemize}[leftmargin=*]
    \item \textbf{The Deterministic Track (Main Stream):} Executes the standard sequence of MoE operators: \textit{Attention}, \textit{All-to-All Dispatch}, \textit{MoE Computation}, and \textit{All-to-All Combine}. While this track remains semantically strictly equivalent to standard execution, its alternating resource demands (compute-bound vs. bandwidth-bound) expose valuable ``execution slacks'' for the auxiliary track.
    
    \item \textbf{The Auxiliary Track (Control Plane):} Operates asynchronously alongside the main stream via three pipelined stages: (1) \textit{Lookahead Prediction} forecasts expert activation for the upcoming layer; (2) \textit{Balance-Optimal Planning} derives the optimal replication strategy; and (3) \textit{Preemptive Expert Transfer} materializes the plan via P2P communication. 
\end{itemize}

We denote by $\mathbf{P}'$ the baseline expert placement (e.g., the static EP sharding placement), and by $\mathbf{P}$ the updated placement after applying the current planning decision.
For expert transfers, $\Delta^{in}_r$ and $\Delta^{out}_r$ represent the sets of experts to be prefetched into and evicted from rank $r$, respectively.
 
\subsection{Predictive Lookahead Mechanism}
\label{subsec:prediction}

To enable proactive balancing, \sysname{} must break the synchronous dependency of standard MoE architectures, where expert activation is known only after the gating network executes. We therefore introduce \textit{Lookahead Gating} to anticipate the next-layer expert distribution one layer ahead.

\paragraph{Gate-Initialized Lookahead Predictor.}
Leveraging the empirical continuity of hidden states across adjacent transformer layers, we construct a lookahead predictor by reusing the target layer's pre-trained router as a strong prior.
Concretely, for each MoE layer $L$, we clone its router parameters $(\mathbf{W}_L, \mathbf{b}_L)$ and freeze them as a base. To compensate for cross-layer feature drift while avoiding cold-start instability, we attach a lightweight, trainable residual MLP:
\begin{equation}
\hat{\mathbf{l}}_L = \underbrace{\mathbf{W}_L \mathbf{h}_{L-1} + {\mathbf{b}}_L}_{\text{Frozen Prior}} + \underbrace{\hat{\mathbf{W}}^{2}_L \, \sigma\!\big(\hat{\mathbf{W}}^{1}_L \mathbf{h}_{L-1}\big)}_{\text{Trainable Residual}},
\end{equation}
where $\mathbf{h}_{L-1}$ is the hidden state from the previous layer and $\sigma(\cdot)$ is the SiLU activation.
We zero-initialize the residual component to match the cloned router initially, ensuring stable starting performance while allowing gradual, data-driven corrections. 
Crucially, local predictions derived from $\hat{\mathbf{l}}_L$ are aggregated via a lightweight All-Gather to enable global prefetch planning; the actual token dispatch strictly follows the ground-truth router outputs during execution, preserving semantic equivalence.

\paragraph{Scale-Driven Online Distillation.}
To ensure robustness against dynamic semantic shifts in real-world traffic, we adopt a scale-driven distillation strategy inspired by Eagle-3~\cite{li2025eagle}. We treat the continuous stream of inference requests spanning diverse domains as a mixed dataset.
By minimizing the Cross-Entropy loss between the predictor's output and the ground-truth router's probability distribution, we force the lightweight MLP to align its trajectory with the actual gating logic.
This massive exposure to online data enables the compact predictor to generalize to complex routing patterns, achieving $\approx$90\% Top-K accuracy while incurring negligible overhead.

\subsection{Balance-Optimal Planning}
\label{subsec:planning}

Given the predicted per-expert workload $\hat{\mathbf{n}}$, the solver jointly optimizes expert placement $\mathbf{P}$ and per-expert token assignment $\mathbf{A}$ after planning.
This formulation minimizes critical path latency while ensuring that the cost of dynamic expert replication is strictly bounded by the computation phase of the concurrent pipeline.

\paragraph{Optimization Formulation.}
We formulate straggler mitigation as a resource allocation problem subject to routing conservation and latency hiding constraints. The objective is to minimize the bottleneck rank's latency:
\begin{equation} 
\begin{aligned}
\min_{\mathbf{P}, \mathbf{A}} \quad & \max_{r} \Big( T_{comp}^r(\mathbf{A}) + T_{comm}^r(\mathbf{A}) \Big) \\
\textrm{s.t.} & \;\;  n_{e,r} > 0 \implies \mathbf{P}_{r,e} = 1, \quad \sum_{r} n_{e,r} = {n}_e, \;\forall e, r \\
& \underbrace{T^{r}_{trans}(\mathbf{P})}_{\text{Prefetch Latency}} \le \underbrace{T_{window}^r}_{\text{Hiding Window}}, \quad \forall r
\end{aligned}
\end{equation}
$T_{window}^r$ here denotes the rank-local overlap window available for expert transfers, i.e., the executation of attention or Grouped GEMM. The first constraint ensures routing validity. The second constraint enforces a zero-penalty, bounding the prefetching latency within the available computation window. This guarantees that transfers are fully overlapped, preventing bandwidth contention with critical communications.

\paragraph{Greedy Rebalancing Strategy.}
Since finding the global optimum for joint placement and routing is computationally expensive, we employ an iterative heuristic detailed in Algorithm~\ref{alg:greedy_solver}.
The process repeatedly identifies the global bottleneck rank $r_{src}$ and pairs it with the least loaded rank $r_{dst}$ to offload the hottest expert $e^*$.
Crucially, every replication move is gated by a \textit{dual-side budget check} (Line 9) to ensure the transfer latency strictly fits within the hiding window.
Upon validation, we apply a \textit{locality-aware water-filling rebalance} strategy to determine the optimal token distribution.
Adhering to a locality-first principle, tokens generated on $r_{src}$ remain pinned to the local replica to eliminate network overhead.
In contrast, remote traffic comprising requests for $e^*$ originating from ranks without local replicas is dynamically partitioned among all available replicas. 
Rather than enforcing strict peer equality, this redistribution follows a water-filling logic at rank-level granularity, greedily redirecting remote tokens to $r_{dst}$ until the load on $r_{src}$ aligns with the cluster-wide average or the transferable pool is exhausted.
The loop persists until convergence or the iteration budget is consumed, ensuring the planning phase completes within the strict lookahead timeframe to yield the optimized expert placement $\mathbf{P}$ and routing assignment $\mathbf{A}$.

\begin{algorithm}[t]
\caption{Greedy Balance-Optimal Planning}
\label{alg:greedy_solver}
\small
\begin{algorithmic}[1]
\Require Predicted Workload $\hat{\mathbf{n}}$, Baseline Placement $\mathbf{P}'$
% , Hiding Budget $T_{window}$
\Ensure Final Placement $\mathbf{P}$, Routing Assignment $\mathbf{A}$
\State Initialize sets $\Delta^{in}_r, \Delta^{out}_r \leftarrow \emptyset$ for all $r$; \;  $k \leftarrow 0$
\State Initialize $\mathbf{A}$ using $\hat{\mathbf{n}}$ and $\mathbf{P}'$ (Locality-First)
\State $\mathbf{L} \leftarrow \text{ComputeLatencies}(\mathbf{A})$ 
\Loop
    \State $r_{src} \leftarrow \arg\max \mathbf{L}$ \Comment{Identify bottleneck rank}
    \State $r_{dst} \leftarrow \arg\min \mathbf{L}$ \Comment{Identify helper rank}
    \State $e^* \leftarrow \text{SelectHeavyExpert}(r_{src}, \hat{\mathbf{n}})$
    
    \If{\textbf{not} \text{CheckDualBudget}($r_{src}, r_{dst}, e^*, \Delta^{in/out}, \mathbf{L}$)} 
        \State Mark pair $(r_{src}, r_{dst})$ invalid; \textbf{continue} 
    \EndIf 
    
    \State $(\mathbf{A}', \text{gain}) \leftarrow \text{WaterFillingRebalance}(e^*, r_{src}, r_{dst}, \mathbf{A})$
    
    \If{$\text{gain} \le \epsilon \; \textbf{or} \; k \ge k_{max}$} 
        \State \textbf{break} \Comment{Converged or budget exhausted}
    \EndIf
    
    \State $\Delta^{out}_{r_{src}} \leftarrow \Delta^{out}_{r_{src}} \cup \{e^*\}; \quad \Delta^{in}_{r_{dst}} \leftarrow \Delta^{in}_{r_{dst}} \cup \{e^*\}$
    \State $\mathbf{A} \leftarrow \mathbf{A}'; \;\; \mathbf{L} \leftarrow \text{ComputeLatencies}(\mathbf{A})$
    \State $k \leftarrow k + 1$
\EndLoop
\State $\mathbf{P} \leftarrow \text{UpdatePlacement}(\Delta^{in/out}, \mathbf{P}')$
\State \textbf{return} $\mathbf{P}, \mathbf{A}$
\vspace{-0.5em}
\end{algorithmic}
\end{algorithm}

\subsection{Phase-Locked Co-Scheduling}
\label{subsec:scheduling}

To materialize the plan without stalling the critical path, \sysname{} implements a Phase-Locked Co-Scheduling policy. This mechanism maps each stage of the auxiliary track to a complementary, orthogonal phase in the main track, ensuring zero contention on hardware resources.

\paragraph{Orthogonal Pipelining with Split-Phase Transmission. }
To neutralize overheads, \sysname{} employs a resource-aware scheduling policy that maps auxiliary-track tasks to complementary main-track phases.
On the compute side, the lightweight MLP-based predictor and the single-SM optimized planning solver initiate concurrently with the bandwidth-bound All-to-All dispatch.
While the predictor is hidden by the dispatch latency, the solver's minimal footprint permits non-intrusive overlap with the subsequent MoE computation.
On the network side, the bandwidth-intensive expert transfer is hidden behind compute-heavy windows via a split-phase transmission mechanism. To prevent contention, transfer initiates during layer $L$'s  MoE computation but is preemptively suspended to yield bandwidth to the critical All-to-All Combine phase; it resumes only after the combine completes, finalizing during layer $L+1$'s attention. This orchestration ensures that management overheads are strictly masked by orthogonal hardware resources.

%% file: sections/5-implementation.tex
\section{Implementation}
\label{sec:impl}

We implement \sysname{} atop the SGLang~\cite{zheng2024sglang} framework, integrating DeepEP~\cite{zhao2025deepep} (normal mode) as the communication backend while maintaining CUDA Graph capture compatibility. We leverage symmetric memory provided by NVSHMEM~\cite{nvshmem} to manage a dedicated replicated-expert buffer region. For prediction, we implement a lightweight global All-Gather with NVSHMEM primitives to synchronize per-rank estimates. For planning, the solver is realized as a single-SM CUDA kernel that performs serial iterative updates, with a hard cap of $k_{max}=16$ iterations to bound overhead. For prefetching, we use a custom Triton~\cite{tillet2019triton} kernel to issue remote put operations with controlled SM occupancy. To support at most three redundant experts per rank, we adopt double buffering for the replica region, limiting memory overhead to six expert slots per device and enabling asynchronous writes of next-layer weights while the current layer computes.

%% file: sections/6-experiment.tex
\section{Experiments}
\label{sec:experiments}

We conduct a comprehensive evaluation to demonstrate the efficiency of \sysname{}. Specifically, our experiments are designed to answer the following key questions:

\textbf{\underline{End-to-End Performance:}} How much acceleration does \sysname{} achieve in both prefill and decoding phases compared to state-of-the-art baselines? (\S\ref{sec:e2e})

\textbf{\underline{Robustness:}} Can \sysname{} maintain stability under dynamic workloads with abrupt semantic shifts? (\S\ref{sec:robustness})

\textbf{\underline{Predictor Fidelity:}} Does the lookahead predictor capture expert activation patterns with sufficient accuracy? (\S\ref{subsec:predictor_analysis})

\textbf{\underline{Latency Breakdown:}}  How effectively does our dual-track pipeline overlap communication with computation to hide system overheads at the micro-operation level? (\S\ref{sec:breakdown})

\subsection{Experimental Setup}
\label{sec:setup}

\paragraph{Environments.}
We evaluate \sysname{} on an 8$\times$NVIDIA Hopper-141GB node interconnected via 900 GB/s NVSwitch. The software stack includes PyTorch 2.9, CUDA 12.9, NCCL 2.27.3~\cite{NCCL} and NVSHMEM 3.3.20~\cite{nvshmem}. 
% We evaluate \sysname{} on an 8$\times$NVIDIA H200-141GB node interconnected via 900 GB/s NVSwitch. The software stack includes PyTorch 2.9, CUDA 12.9, NCCL 2.27.3~\cite{NCCL} and NVSHMEM 3.3.20~\cite{nvshmem}. 
% We plan to extend our evaluation to bandwidth-constrained settings (e.g., H800 nodes or cross-node clusters) in subsequent updates.

\paragraph{Models and Datasets.}
We benchmark on two models representing distinct sparsity configurations: \textbf{Qwen3-MoE-235B} (128 experts, Top-8, 93 layers, BF16) and the more sparse \textbf{GPT-OSS-120B} (128 experts, Top-4, 36 layers, BF16).
For evaluation, we construct three datasets: \textit{Chinese} and \textit{Code}, aggregated from multiple open-source corpora (e.g., Alpaca-zh~\cite{peng2023instruction}, CodeAlpaca-20k~\cite{chaudhary2023code}, OpenAI-humaneval~\cite{chen2021evaluating}), and a synthetic \textit{Repeat} dataset. The latter is constructed by duplicating a narrow set of prompts to simulate extreme expert skew within the $ep$=8 environment.

\paragraph{Baselines.}
We compare \sysname{} against two representative systems: \textbf{SGLang}~\cite{zheng2024sglang}, the standard EP baseline employing static sharded placement; and \textbf{DeepSeek-EPLB}~\cite{zhao2025deepep}, a statistic-based load balancing strategy. For EPLB, we configure 2 redundant expert slots per layer per rank, constraining the global rebalancing transfer to complete within 2 decoding steps.

\subsection{End-to-End Performance}
\label{sec:e2e}

\begin{figure}[t]
    \centering
    \includegraphics[width=\linewidth]{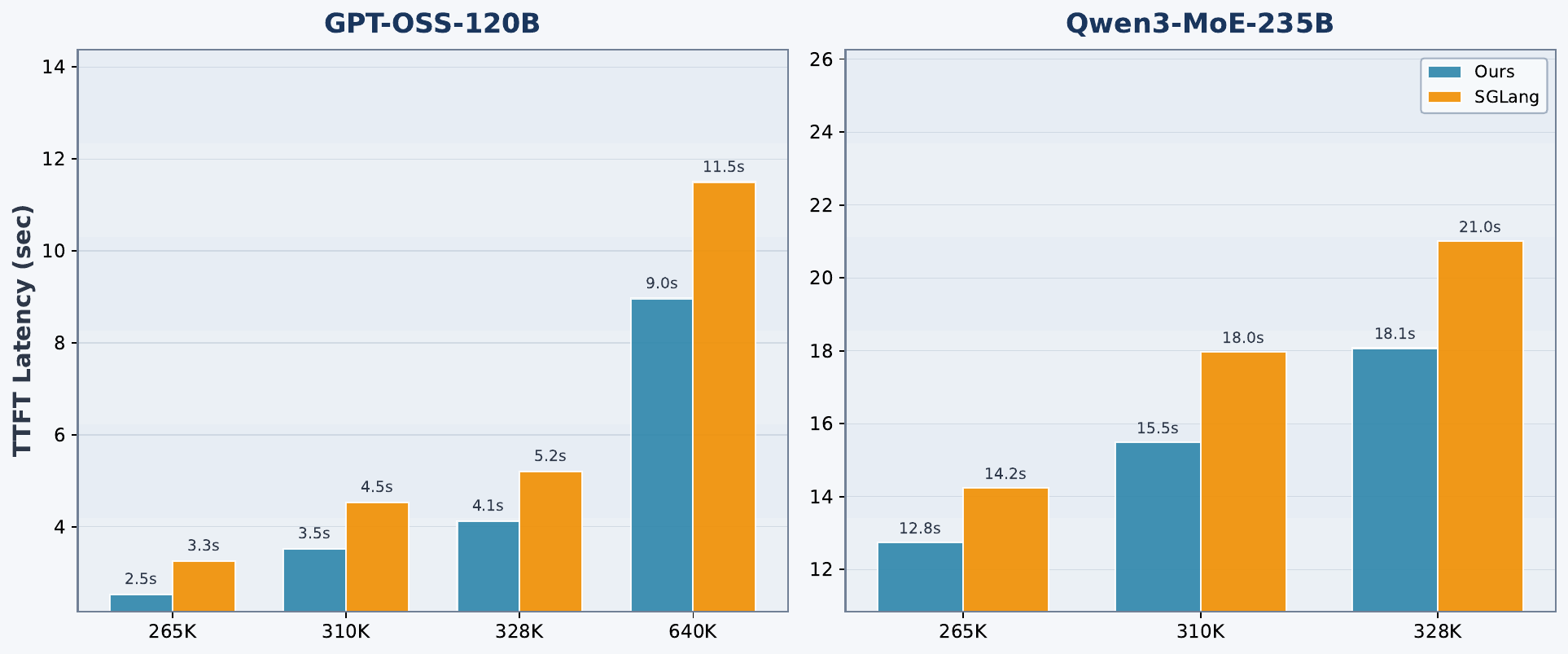} 
     \caption{\textbf{Prefill latency scaling.} Evaluated on an node with $ep=8$. We use chunked prefill of 8K (GPT-OSS) / 16K (Qwen3-MoE) tokens per rank; the x-axis shows total input tokens across ranks. We omit DeepSeek-EPLB because extra replicas can trigger OOM under prefill memory pressure, and reactive transfers are too costly given the limited prefill steps.}
     \vspace{-1em}
    \label{fig:ttft}
\end{figure}

\paragraph{Prefill Latency (TTFT).}
We first evaluate the performance during the compute-intensive prefill phase, as detailed in Figure~\ref{fig:ttft}.
Standard EP implementation suffers significantly from the bursty nature of prompt processing. \sysname{} effectively neutralizes these stragglers. 
By dynamically balancing the expert load, \sysname{} achieves consistent acceleration across both models and total input tokens, peaking at a \textbf{1.32$\times$} speedup compared to SGLang. 
Although the hybrid parallelism (DP for attention module) introduces potential attention workload skew, its impact is mitigated in our experiments by the usage of chunked prefill and dominance of short-context prompts, leaving MoE stragglers as the predominant bottleneck.
Furthermore, these gains are more pronounced on the sparser GPT-OSS-120B model, as its inherently higher $\mathcal{IR}$ exacerbates straggler bottlenecks, thereby offering a larger optimization margin compared to Qwen3-MoE.
Notably, we exclude DeepSeek-EPLB from this evaluation due to its fundamental incompatibility with the prefill regime.
First, its reliance on historical statistics proves ineffective against the rapid, instantaneous semantic shift characteristic of the condensed prefill stage.
Second, unlike \sysname{}'s hidden scheduling, EPLB's rebalancing incurs transfer overheads that significantly outweigh the potential gains given the limited number of prefill steps.
Finally, the alternative of static expert replication is infeasible, as the additional memory footprint triggers OOM errors under the high memory pressure of large-batch processing.

\begin{figure}[t]
    \centering
    \includegraphics[width=0.88\linewidth]{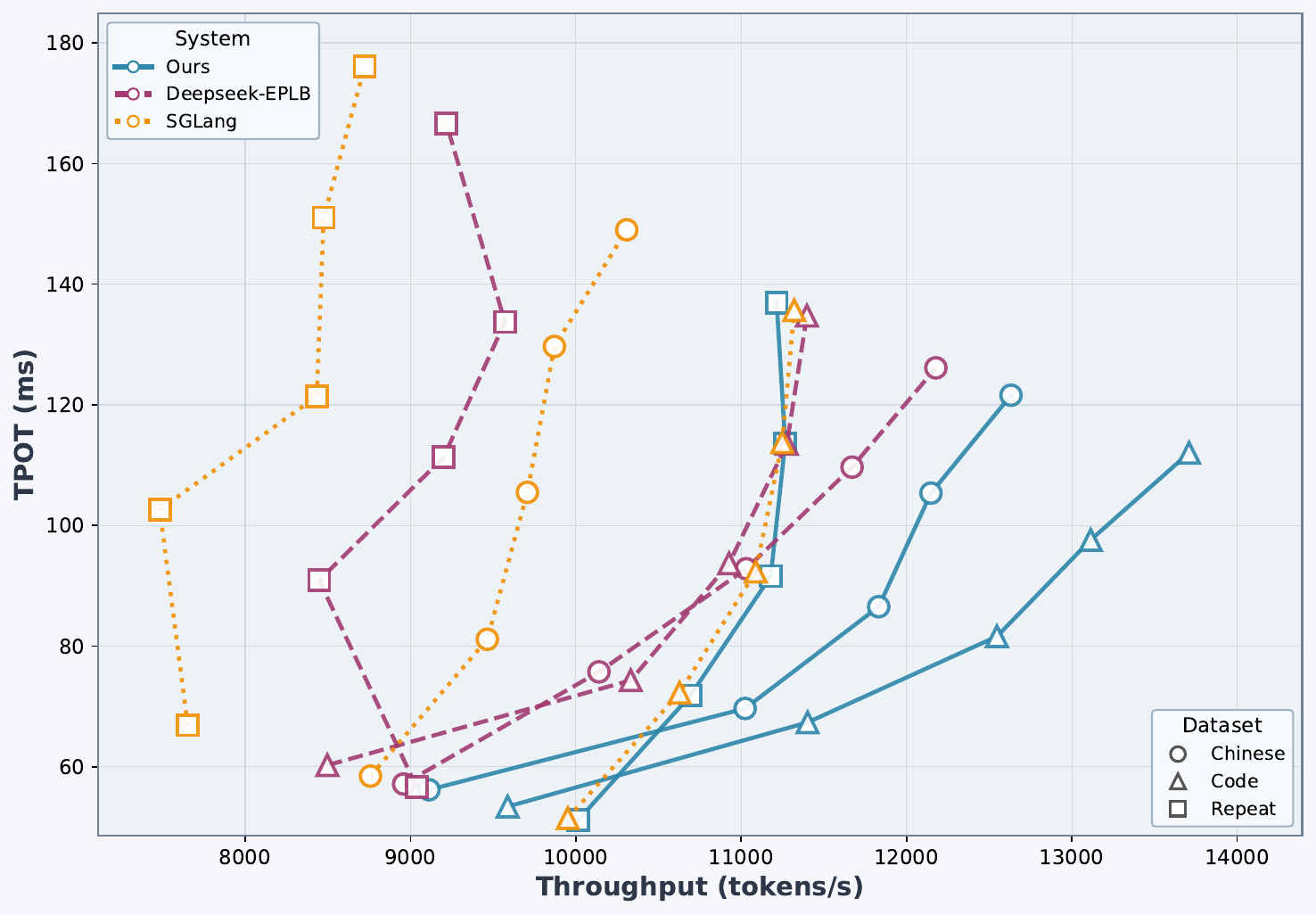} 
    \caption{\textbf{Throughput-latency Pareto frontier of decoding stage.} GPT-OSS with $ep=8$. We sweep per-rank batch size from 512 to 1536 on \textit{Chinese}, \textit{Code}, and \textit{Repeat}. \sysname{} consistently dominates the frontier; on \textit{Repeat}, it avoids the large latency spikes caused by extreme expert skewness in baselines.}
    \label{fig:tpot_tradeoff}
    \vspace{-1em}
\end{figure}

\paragraph{Decoding Throughput-Latency Trade-off.}
Figure~\ref{fig:tpot_tradeoff} illustrates the system performance during the decoding phase. We report the average throughput over the initial 500 decoding steps by sweeping batch sizes.
\sysname{} consistently pushes the Pareto frontier towards the optimal bottom-right corner across all datasets, demonstrating a superior trade-off between throughput and latency.
Compared to DeepSeek-EPLB configured with one-shot rebalancing, \sysname{} achieves up to \textbf{1.26$\times$} higher throughput at the same batch size. The performance gap stems from the temporal dynamics of the workload: EPLB's static placement, derived from a single snapshot, progressively degrades as the semantic distribution drifts over the 500-step window. In contrast, \sysname{}'s \textit{continuous lookahead prediction} ensures optimal expert locality for every layer of every step, thereby providing robustness against such volatility, which is especially evident on the high-skew \textit{Repeat} dataset. 
Furthermore, \sysname{} optimizes memory efficiency by cyclically reusing expert slots, avoiding EPLB's requirement for static per-layer placeholders that that compete with KV cache for memory. Our dynamic approach strictly limits the redundancy footprint, preserving maximum capacity for long-context inference.

\subsection{Robustness to Semantic Shifts}
\label{sec:robustness}

\begin{figure}[h]
    \centering
    \includegraphics[width=0.98\linewidth]{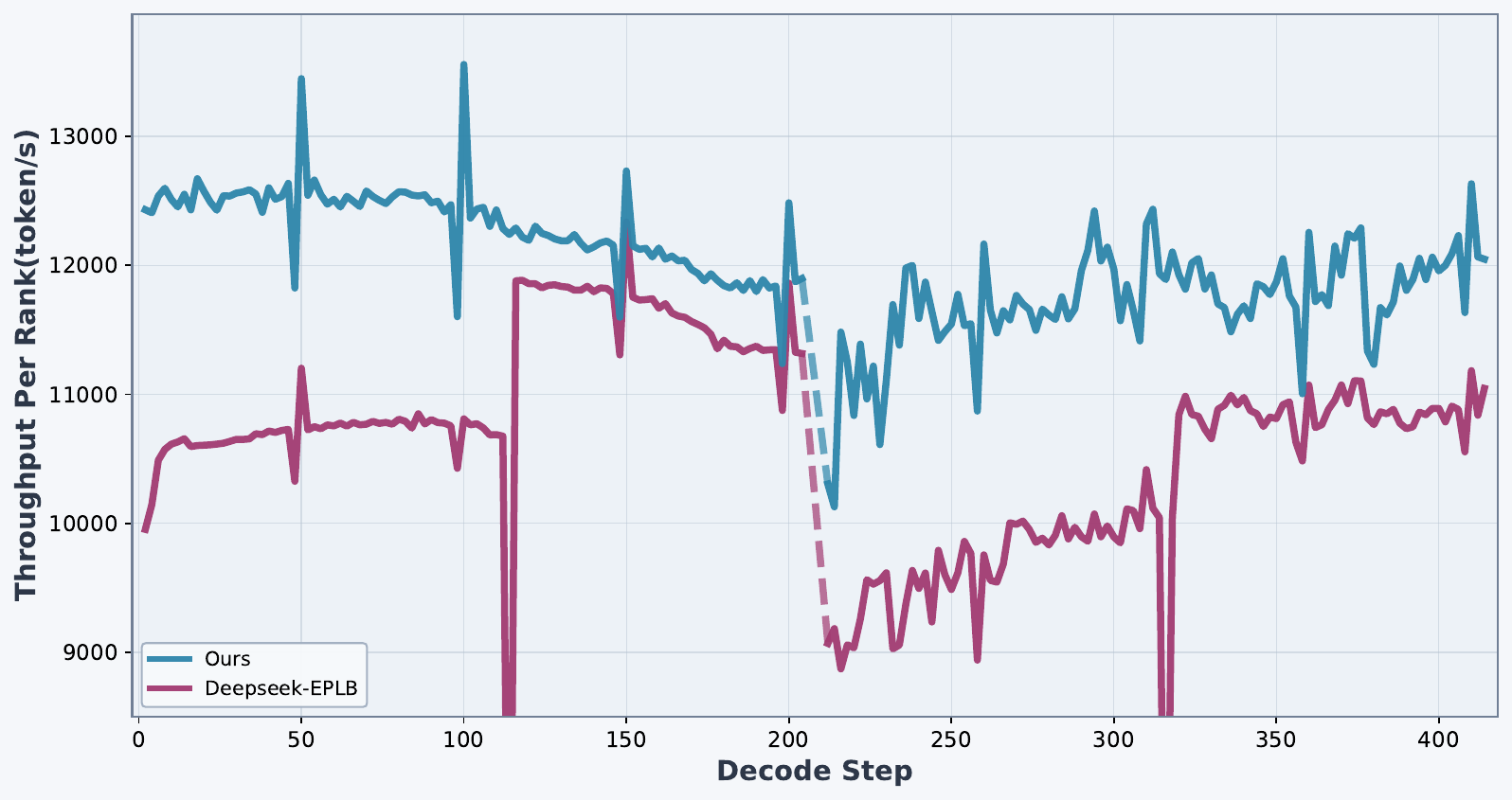} 
     \caption{\textbf{Throughput under abrupt semantic shifts.} GPT-OSS-120B with $ep=8$: the workload switches from \textit{Code} to \textit{Chinese} at step $\approx$200. DeepSeek-EPLB incurs warm-up and degrades after the shift due to stale placement, while \sysname{} remains stable via real-time predictive planning.}
    \label{fig:robustness}
    \vspace{-1em}
\end{figure}

To validate system robustness against workload volatility, we designed a ``stress test" to simulate abrupt semantic transitions. As shown in Figure~\ref{fig:robustness}, we initiate the decoding process with the \textit{Code} dataset, followed by an instantaneous switch to the \textit{Chinese} dataset with higher $\mathcal{IR}$ at step $\approx$200.
DeepSeek-EPLB exposes the limitation of historical statistic-based approaches.
Initially, EPLB operates with a default placement with no redundant expert, suffering from suboptimal throughput until step $\approx$110. At this point, sufficient historical activation data is collected to trigger a rebalancing event, resulting in a visible performance jump.
However, this gain is temporary, when the workload shifts at step$\approx$205, EPLB's performance drastically degrades. This occurs because the system retains the expert placement optimized for the previous distribution, which is now mismatched with the new high-skew workload. 
% EPLB remains trapped in this ``stale state", exhibiting severe throughput fluctuations until sufficient new statistics are gathered to trigger a second rebalancing operation.
In sharp contrast, \sysname{} demonstrates robustness to volatility. 
By leveraging the lookahead predictor, \sysname{} anticipates the expert hotspots of the next layer in real time, rather than relying on the statistics of past steps.
Consequently, \sysname{} requires no warm-up period and maintains a stable, high-throughput trajectory across the shift boundary. The system instantly adapts to the new dataset without the lag observed in EPLB, proving its suitability for environments where request variability is the norm.

\begin{figure}[th]
    \centering
    \includegraphics[width=\linewidth]{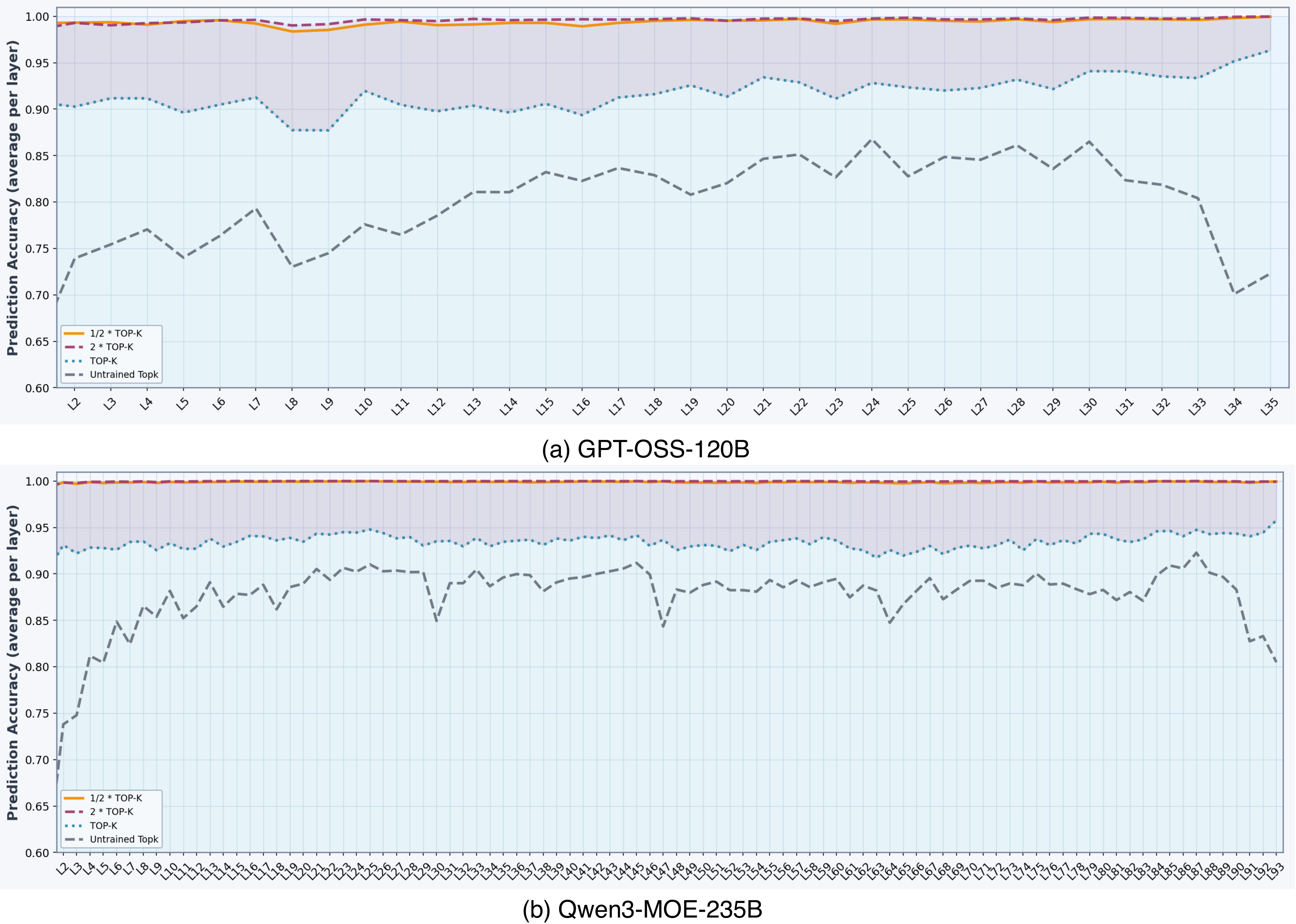} 
    \caption{\textbf{Predictor fidelity across layers.} The untrained baseline (frozen router only) suffers from feature drift, while online distillation improves Top-$K$ accuracy to $\approx 90\%$. The Top-Half-$K$ hit rate and 2$\times$Top-$K$ recall (within a 2$\times$ prediction window) both approach 100\%.}
    \label{fig:prediction_heatmap}
\end{figure}

\subsection{Analysis of Predictor Fidelity}
\label{subsec:predictor_analysis}

The efficacy of \sysname{}'s pipelining depends on the predictor's ability to anticipate expert activation with high fidelity. We first validate the necessity of the trainable residual component: as shown in the Figure~\ref{fig:prediction_heatmap}, while the \textit{Untrained} baseline suffers from feature drift with only around 70\%--80\% accuracy, our online distillation strategy significantly corrects this, elevating the \textit{Top-K Accuracy} to \textbf{87\%--94\%} across layers and ensuring bandwidth is consumed only for valid transfers. Beyond standard accuracy, the predictor exhibits near-deterministic reliability for critical workload components; the \textit{Top-Half-K Hit-Rate} and the \textit{2$\times$Top-K Recall} consistently approach \textbf{100\%}, serving as a virtually perfect predictor against false negatives. Crucially, this certainty implies that the potential of predictive execution extends far beyond expert prefetching, especially in disaggregation frameworks~\cite{zhu2025megascale,wang2025step}: since the destinations of the majority of tokens are known effectively before routing computation completes, future optimizations could leverage this signal to pre-dispatch hidden states to high-confidence experts, potentially overlapping the entire All-to-All communication latency with routing itself.

\begin{figure}[t]
    \centering
    \includegraphics[width=0.97\linewidth]{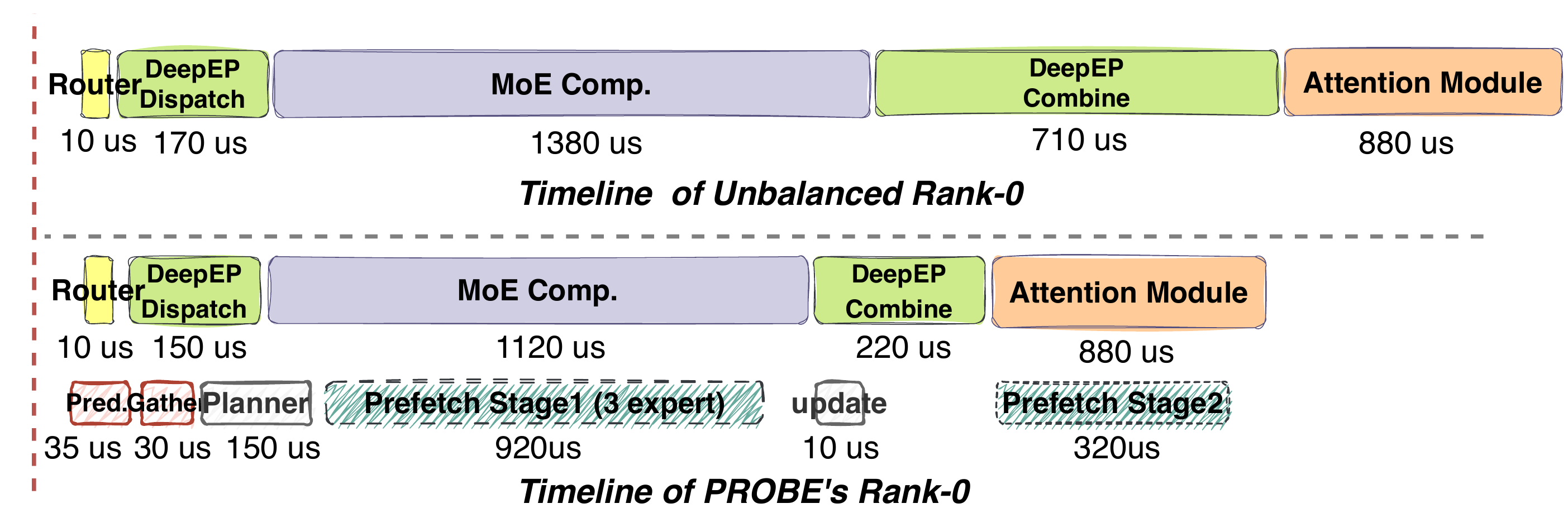} 
    \caption{\textbf{Timeline breakdown of an single decoding step.} GPT-OSS with $ep=8$ and per-rank $b=768$, averaged over Layers 1--35 (excluding Layer 0). \textbf{Top:} baseline Rank-0 timeline, where \textit{Combine} is inflated by synchronization with stragglers. \textbf{Bottom:} \sysname{}'s dual-track timeline; \textit{Prefetch} visualizes transferring up to three experts (0--3 selected per layer) to demonstrate hiding.}
    \vspace{-1em}
    \label{fig:breakdown}
\end{figure}

\subsection{Latency Breakdown}
\label{sec:breakdown}

Figure~\ref{fig:breakdown} validates \sysname{}'s dual-track scheduling via the averaged micro-operation timeline across Layers 1-35 of GPT-OSS. The breakdown confirms the effective concealment of all control overheads: the \textbf{Predict} phase (comprising MLP inference and global Gather) and the single-SM \textbf{Planner} execute concurrently with \textit{Dispatch}, with the planner's tail latency naturally overlapped by the  \textit{MoE Compute}. Furthermore, the \textbf{Prefetch} latency with 3 expert budget is masked via split-phase transmission, overlapping sequentially with the current \textit{MoE Compute} and the subsequent layer's \textit{Attention}, while the \textbf{Update} phase prepares expert and assignment masks without stalling execution. 
Crucially, \sysname{} effectively neutralizes stragglers: the average $\mathcal{IR}$ across 35 layers sees a modest reduction from \textbf{2.13} to \textbf{1.09}, accompanied by a substantial drop in computation latency skew (Max/Avg) from \textbf{2.27} to \textbf{1.18}. This alignment eliminates synchronization idle time, which otherwise manifests as inflated \textit{Combine} latency. Notably, given the abundant bandwidth, the visible reduction in the \textit{Combine} phase is driven primarily by this elimination of wait times rather than accelerated data transfer.

%% file: sections/7-conclusion.tex
\section{Conclusion}
\label{sec:conclusion}

In this paper, we identify that MoE inference efficiency is fundamentally limited by the interplay of spatial stragglers and temporal workload volatility.
To address these challenges, we propose \sysname{}, an inference system that shifts balancing paradigm from reactive adjustment to proactive preparation.
\sysname{} leverages Continuous Lookahead Pipelining that combining a high-fidelity predictor, a hardware-aware planner, and phase-locked co-scheduling, to co-balance computation and communication in real time without stalling execution.
Extensive evaluations show that \sysname{} improves both prefill latency and decoding throughput over state-of-the-art baselines.
Collectively, our results validate predictive lookahead execution as a promising approach for efficient trillion-parameter MoE deployment.